# On the enhancement of non-functional requirements for cloud-assisted middleware-based IoT and other applications

Samir Medjiah [1,2] and Christophe Chassot [1,3]

[1] CNRS, LAAS, 7 avenue du Colonel Roche, F-31400 Toulouse, France
[2] Univ. Toulouse, UPS, LAAS, F-31400 Toulouse, France
[3] Univ. Toulouse, INSA, LAAS, F-31400 Toulouse, France

`{medjiah, chassot}@laas.fr`

**Abstract.** The Internet of objects (IoT) will have to meet the non-functional needs (QoS, security, etc.) of new business applications supported by the cloud. To do this, the interactions between the underlying application software and the communicating objects will rely on networks and communication middleware with configurable, programmable and dynamically deployable capabilities. These capabilities will be available both on pre-existing entities but also on virtual entities, i.e. that will be dynamically created in the Cloud according to the need. In this new ecosystem, meeting the end-to-end QoS needs of these future applications is a major challenge. This challenge has particularly to be tackled both at the level of the Middleware intermediary entities and at the level of the networks interconnecting these entities. In this context, this paper presents our approach for a self-adaptive QoS management at the middleware level for IoT applications. This approach is aimed at: 1) taking advantage of the technological opportunities offered by the Cloud, the dynamic deployment of processing functions and the autonomic computing paradigm, 2) taking into account the heterogeneity of the solutions that will coexist in this landscape, and 3) ensuring the consistency of the (re) configuration choices thanks to appropriate theoretical tools.

**Keywords:** Cloud-assisted, middleware, micro-service, service oriented, QoS management, IoT, autonomic computing, dynamic (re)configuration, dematerialized network function, scalability, NFV, SDN, model-driven,

## 1 Introduction

### 1.1 Context

The future Internet, including not only usual terminals but more generally any form of communicating objects, will have to meet non-functional needs (security and quality of service - QoS in particular) of new business applications, taking advantage of Cloud in various domains such as remote supervision, personal assistance, and urban transport. To do this, the interactions between the application software and these objects will be based on communication networks and "middleware" layer allowing both to abstract applications from the complexity of underlying technologies (networks and connected objects), but also to avoid "vertical" fragmentation of IoT solutions. In this new ecosystem, several challenges already addressed in the "classical" Internet are to be (re)-considered in particular the performance of the whole communication system (i.e. including network layers and middleware) in response to the expressed Application-level QoS requirements, for instance, in terms of high service availability or bounded response time. From 2010, a major standardization effort has been conducted at the Middleware level, notably via ETSI and the oneM2M consortium [1][2]. However, the proposed solutions remain a point of fragility in front of the expected performances. Some works begin to address the QoS problem at the IoT Middleware level by proposing mechanisms for managing application level traffic (for example, delaying less priority requests in case of congestion) and / or allocated resources of the underlying machines [3].

In this context, the advent of the so-called "virtualization" technologies initially linked to the Cloud, makes it now possible to consider the deployment of these mechanisms not only on dedicated equipment (typically a Gateway in the world of IoT) but in private or public data centers with hypervisors offering the required functional capabilities. The concept of "virtual network function" - VNF ("virtual" is not necessarily in the sense of "implemented" on dedicated equipment) has been defined by the ETSI as part of its work on standardization of the NFV technology [4]. This concept is today to be considered in a wider framework involving the deployment of such functions on any node that can host and execute the corresponding program, whether it has a hypervisor or not (i.e. serverless paradigm [5]). It is for example possible to deploy and launch an executable applicative program on a simple PC without interrupting the execution of its operating system. It is also possible to deploy an application module and integrate it dynamically within an application code whose design is based on a component or micro services-oriented approach.

This analysis leads us to (re)-define the concept of "network function" (NF) (cf. Figure 1), which basically consists in a given processing of packets of any level (Application, Transport, Network, ...). In our vision, the concept of NF integrates and extends the concept of ETSI VNF which appears as a special case of what we call a "dematerialized" (DNF) network function, i.e. deployed outside of its original environment (as opposed to the "physical" network functions). A DNF can then be defined either as a VNF or as an application network function (ANF) consisting in an executable applicative program or an application module.

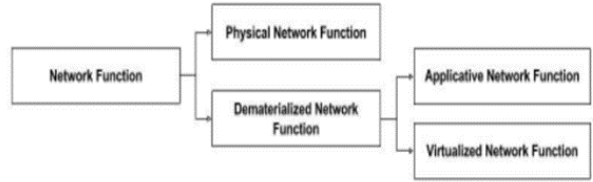

**Fig. 1.** Network function concept.

**1.2 Problem statement**

Let us then consider a vision of an IoT underpinned by a generic IT infrastructure including not only dedicated middleware and network entities (typically gateway / server and routers) but also private and public data centers, and also simple machines on which it is possible to deploy code in the form of an executable application and / or an application module. It should be noted that this code can be applied at the Application level (in the OSI sense of the term), but also at the Transport level (i.e. TCP level of the Internet stack) if the latter is built according to a component-based design [6][7]. In front of the performance problems, the global issue we consider in this positioning article is to define how to take advantage of the characteristics (opportunities and constraints) of the available IT infrastructure with the aim to implement a dynamic management approach of QoS-oriented mechanisms meeting the application requirements, at different levels of the communication system, typically Middleware, Transport and/or Network levels.

**1.3 Approach**

Our approach consists in designing, developing and testing generic architecture models for a self-adaptive management of QoS-oriented network functions (in the NF sense) at the different levels of the communication system:
- taking advantage of the technological opportunities associated with the dynamic deployment of network functions, but also of software-driven networks (in the SDN sense),
- taking into account the factual heterogeneity of the solutions being deployed,
- ensuring the consistency of the configuration and reconfiguration choices made for each level through appropriate theoretical tools.

In our work, the autonomy management is based on the autonomic computing paradigm (IBM). In this paper, we mainly focus on the Middleware level.

The rest of this paper is structured as follows. We first detail in section II the notion of middleware-based IoT application; we also position the Cloud and its opportunities. Our approach for a dynamic and autonomous management of network functions (here

QoS oriented) is presented in Section III. The key challenges associated with this approach are detailed in Section IV. The conclusions and perspectives of the paper are finally presented in Section V.

## 2   Definition and motivation of "cloud-assisted middleware-based applications"

Following the classic deployment of IoT solutions where we have smart objects such a sensors or actuators connected to application servers through one or multiple gateways, communications' management can be very tricky. Indeed, despite the required huge number of connected things to fulfill the application objectives, these devices often use different networking technologies, communication protocols and data formats, making the overall communication between the application logic and the physical objects difficult to achieve. In this context, an application can greatly benefit from the services of software middleware that abstract the application from the underlying networking specifics. Thus, through an IoT middleware, the application logic uses simple APIs that most often follow a RESTful model and use universal protocols such as HTTP or CoAP to interact with the physical objects in order to query sensed data or to trigger object's commands. In this case, the application developer is not required to implement or even to learn specific IoT protocols such as ZigBee, Z-Wave, Bluetooth, etc. in order to communicate with the physical objects.

For applications' backend, Cloud based deployment has allowed to tackle the non-functional requirements (on particular QoS requirements) through the offered elasticity of resources (storage & compute) being made available to the applications' backend. For example, a virtual machine hosting a MW entity can see its memory (RAM) dynamically resized and scaled up (Vertical Scalability) in order to meet the increasing processing load (e.g. Increase in number of messages routed through this MW entity). Another typical example of the cloudification benefits is the increase in the number of VMs hosting the same MW entity with a load balancing mechanism in order to respond to a sudden load increase.

Cloud based deployment can also allow to tackle functional requirements for example through the dynamic deployment of data/packets processing functions that are implemented in software and previously packaged in virtualization containers (VMs or system-level containers: LXC, Docker, etc.). For example, an application may need its data format to be translated from a text format (e.g. JSON) to a binary format (e.g. BSON) before being exchanged by the MW in order to optimize the data transfer, and translated back again to be received by the final destination. Another example is the deployment of message aggregation function between the application and the MW in order to adapt the MW communications to the underlying networks (e.g. Satellite networks with big data frames).

## 3 Enhancement approach

In this section, we present the typical deployment infrastructure of a Cloud-assisted and middleware-based IoT solution (cf. Figure 2). In this context, the IoT application is accessed using a user terminal such as computers or smartphones. An IoT middleware is linking this application to the physical sensors and actuators. The application backend as well as some MW entities are being deployed in public or private Cloud. Some MW entities may be deployed within the Service Provider private infrastructure using dedicated nodes (typically for IoT gateways) or generic deployment nodes such as COTS hardware.

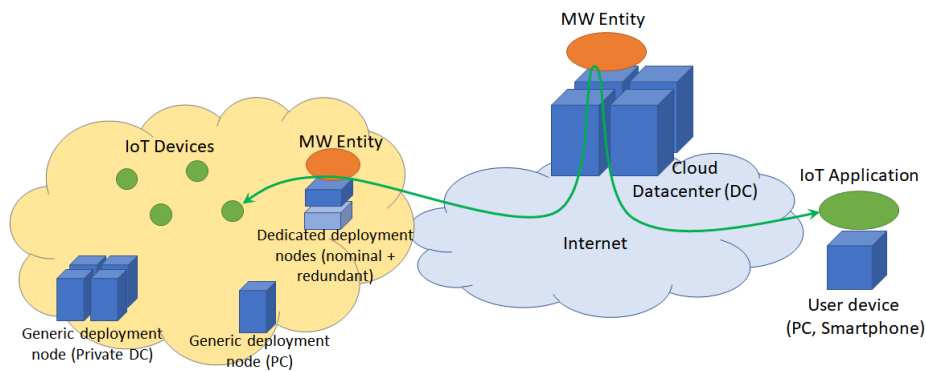

**Fig. 2.** Typical deployment infrastructure of Cloud-assisted middleware-based IoT solutions.

In this context, we consider the following actions strategies:

1. *Vertical Scalability*: as illustrated earlier, this strategy aims to dynamically scale a virtualization container's resources such as memory, storage or compute resource, whether this CV is deployed on a public or private Cloud.
2. *Horizontal Scalability*: this strategy can be implemented through multiple ways:
   a. Activation of backup MW entities already deployed within dedicated nodes in order to achieve redundancy.
   b. Deployment of new instances of CVs hosting MW entities within generic nodes; COTS nodes or private/public Cloud.
3. **Dynamic Function Deployment**: with this strategy, processing function of data/packets at different levels (application, transport or network) can be deployed in specific locations in order to perform message-oriented QoS adaptation actions. Such processing can be deployed through:
   a. an application module (ANF) to change/augment the behavior of a MW entity.
   b. A CV hosting different functions within generic nodes or data centers (public or private)

   These functions can be classified into data-oriented and packet-oriented functions:

   o *Packet*-oriented functions:

- *Classifier*: a function responsible for identifying packet's class in order to achieve differentiated processing.
- *Marker*: a function closely related to the classifier function that is responsible for marking the packets with the appropriate class/category.
- *Scheduler*: a function responsible for packets reordering based on specific scheduling algorithm. This function can rely on the Classier/Marker functions in order to schedule the marked packets based on their marking.
- *Discarder*: a function for dropping marked packets or based on a discarding algorithm
- *Delayer*: a function for delaying marked packets for delaying or based on a specific algorithm.
- …
  - *Data*-oriented functions, examples include:
    - *Transcoder*: a function for translating the payload of packets from one format to another for transfer optimization or interoperability purposes. For example, JSON from/to BSON (optimization), XML from/to JSON (interoperability), etc.
    - *(De)Compressor:* a function for compressing/decompressing the data packets payloads in order to optimize the data transfer especially in constrained networks.
    - …

Relying on these three action strategies, the system is offered a large panel of mechanisms which are different in behavior, requirements and results. Therefore, system reconfiguration can be achieved through the use of one or a combination of all the possible mechanisms. It is then up to the pacification process to choose the appropriate set of mechanisms and their orchestration in order to meet the targeted non-functional requirements and in particular QoS requirements.

Since, these mechanisms are inherently dynamic, autonomous behavior that is based on these mechanisms can be achieved through the implementation of an autonomic manager. In our approach, we have considered the autonomic computing paradigm proposed by IBM [8].

## 4    Key challenges

Several challenges have to be addressed in order to carry out the proposed approach. We present in this section the main challenge we have considered.

**Design of the management system architecture.**
The first challenge is to design the structural and behavioral architecture of the management system leading to the dynamic and autonomous deployment of QoS-oriented network functions. Three areas of research are particularly relevant.

- The first one deals with the OSI level of actions to consider, in our case: Middleware, Transport and Network levels. For each level, knowing the available functional and non-functional constraints and capabilities is a necessary prerequisite. Beyond this, for a multi-level management, the coherence of the adaptation choices is a major issue to avoid overreactions in case of QoS degradation. At the Transport level, the actions consist in adapting the protocol architecture in terms of mechanisms to be implemented in order, on the one hand, to meet the services required by the application, and on the other hand, to be the most appropriate with regards to the features of the underlying network. At the Network level, the SDN technology opens the possibility of configuring the network, for example in dynamically establishing secure and / or QoS-oriented tunnels between two edge routers.
- The second research issue is related to the spectrum covered by the targeted autonomous management. In a distributed middleware context, for instance, the design of an autonomous manager acting on all the servers and gateways involved on the data path is a first possibility. However, due to the number of metrics to be processed but also the communication latency between the managed entities and their manager, it is conceivable that such a solution has scalability limitations. The choice of an architecture based on hierarchical policies is conceivable to meet this limit. For a distributed middleware, we can for example imagine two levels of autonomous management, one local to each middleware entity leading to the adaptation of the behavior and / or resources of each of these entities, a higher level global one to define the QoS objectives to be reached by each local manager.
- The third challenge is related to the implementation choice of the autonomic manager that can be centralized or distributed. Another implementation choice comes from the use (or not) of the virtually infinite resources of the Cloud with the aim to avoid that the manager becomes itself a bottleneck with respect to QoS.

**Models to drive the different steps of the autonomous process**

Several paradigms can be envisaged to implement the logic of the autonomous manager, among them the autonomic computing (AC) model proposed by IBM [8]. In addition to its four objectives, which are self-configuration, self-optimization, self-repair and self-protection, the AC relies on the implementation of a four-step loop dealing with: (1) the Monitoring of the resources and / or the performance of the managed system in order to raise symptoms representing for instance a QoS degradation, (2) the Analysis of the causes explaining the identified failures, (3) the Planning of the adaptation actions to be undertaken, and finally (4) the Execution of these actions by the managed system. The implementation of this so-called MAPE-K loop is based on a Knowledge base including the relevant elements (model information, etc.).

Two main approaches can be envisaged to carry out the Monitoring, Analysis and Planning steps: either by working on the managed system itself (for example a middleware entity) including the traffic to be transferred, or on the basis of models, for example of the managed entities and or the incoming traffic. The first approach is costly in terms of resources and / or overhead. The second approach reduces this extra cost at the possible price of an approximation related to the manipulation of models rather than real elements. It has also as advantage to facilitate both a proactive and a reactive adaptation approach. It is also very useful (and sometimes necessary) to tackle the complexity associated to a multi-level management and to deal with the heterogeneity of the current IT infrastructures in terms of usable technologies. Let us finally note that this e-second approach also allows to ensuring the verification of required properties, such the respect of a given structural functioning schema [9]. The elaboration of such models to drive the different steps of the autonomous process is a challenge in itself that may be addressed thanks to several formalisms and theories such as graphs, queuing theory, etc.

## 5 Conclusion and future work

Within the future virtualized Internet of Things, meeting the end-to-end QoS needs of new cloud-based and other applications is a major challenge. This challenge has particularly to be tackled both at the level of the Middleware intermediary entities and at the level of the networks interconnecting these entities. This paper has presented our approach for a self-adaptive QoS management at the middleware (and other) levels for IoT applications within this new ecosystem. The key challenges associated with this approach have been introduced. Our current and future works deal with the proposal of architectural and model-based solutions towards the identified issues.